\documentstyle[12pt]{article}
\newcommand{\newsection}[1]{
\vspace{15mm}
\pagebreak[3]
\addtocounter{section}{1}
\setcounter{equation}{0}
\setcounter{subsection}{0}
\setcounter{footnote}{0}
\setcounter{fignum}{0}
\begin{center}
{\sc \thesection. #1}
\end{center}
\nopagebreak
\medskip
\nopagebreak}
\newcommand{\newsubsection}[1]{
\vspace{1cm}
\pagebreak[3]
\addtocounter{subsection}{1}
\addcontentsline{toc}{subsection}{\protect
\numberline{\arabic{section}.\arabic{subsection}}{#1}}
\noindent{ \it \thesubsection. #1}                 
\nopagebreak
\vspace{2mm}
\nopagebreak}

\newcommand{\bye}{\end{document}}
\newcommand{\bq}{\begin{quote}}
\newcommand{\eq}{\end{quote}}

\renewcommand{\theequation}{\thesection.\arabic{equation}}

\newcommand{\ben}{\begin{enumerate}}
\newcommand{\een}{\end{enumerate}}
\newlength{\extraspace}
\newlength{\extraspaces}
\newcounter{dummy}
\newcommand{\bc}{\begin{center}}
\newcommand{\ec}{\end{center}}
\newcommand{\be}{\begin{equation}
\addtolength{\abovedisplayskip}{\extraspaces}
\addtolength{\belowdisplayskip}{\extraspaces}
\addtolength{\abovedisplayshortskip}{\extraspace}
\addtolength{\belowdisplayshortskip}{\extraspace}}
\newcommand{\ee}{\end{equation}}

\newcommand{\ba}{\begin{eqnarray}
\addtolength{\abovedisplayskip}{\extraspaces}
\addtolength{\belowdisplayskip}{\extraspaces}
\addtolength{\abovedisplayshortskip}{\extraspace}
\addtolength{\belowdisplayshortskip}{\extraspace}}
\newcommand{\ea}{\end{eqnarray}}
\newcommand{\nonu}{\nonumber \\[.5mm]}
\newcommand{\is}{& \!\! = \!\! &}
\newcommand{\ban}{\begin{eqnarray*}
\addtolength{\abovedisplayskip}{\extraspaces}
\addtolength{\belowdisplayskip}{\extraspaces}
\addtolength{\abovedisplayshortskip}{\extraspace}
\addtolength{\belowdisplayshortskip}{\extraspace}}
\newcommand{\ean}{\end{eqnarray*}}
\newcommand{\baa}{                         
\addtocounter{equation}{1}
\setcounter{dummy}{\value{equation}}
\setcounter{equation}{0}
\renewcommand{\theequation}{\thesection.\arabic{dummy}\alph{equation}}
\begin{eqnarray}
\addtolength{\abovedisplayskip}{\extraspaces}
\addtolength{\belowdisplayskip}{\extraspaces}
\addtolength{\abovedisplayshortskip}{\extraspace}
\addtolength{\belowdisplayshortskip}{\extraspace}}
\newcommand{\eaa}{                                       
\end{eqnarray}
\setcounter{equation}{\value{dummy}}
\renewcommand{\theequation}{\thesection.\arabic{equation}}}

\input epsf
\newcounter{fignum}

\newcounter{tabnum}
\newcounter{xxx}
\newcommand{\bl}{\begin{list}{({\it\roman{xxx}})}{\usecounter{xxx}}}
\newcommand{\el}{\end{list}}
\renewcommand{\d}{{{\partial}}}
\newcommand{\pp}[1]{{\partial \over \partial #1}}             
             
\newcommand{\ppt}[1]{{\partial \over \partial t}}            
\newcommand{\ppx}[1]{{\partial \over \partial x}}            
\newcommand{\pqt}[1]{{\partial^2 \over \partial t^2}}            
\newcommand{\pqx}[1]{{\partial^2  \over \partial x^2}}            
\newcommand{\Pp}[2]{{\partial #1 \over \partial #2}}

\newcommand{\twomatrixd}[4]{{\left(\begin{array}{cc}
\displaystyle #1 & \displaystyle #2\\[2mm]
\displaystyle  #3  & \displaystyle #4 \end{array}\right)}}
\newcommand{\ie}{{\it i.e.\ }}

\newcommand{\eg}{{\it e.g.\ }}

\renewcommand{\l}{\langle}

\renewcommand{\.}{\cdot}
\renewcommand{\ll}{{\lambda}}

\newcommand{\half}{{\textstyle{1\over 2}}}

\newcommand{\Z}{{\Bbb Z}}
\newcommand{\R}{{\Bbb R}}

\newcommand{\C}{{\Bbb C}}

\renewcommand{\H}{{\Bbb H}}

\newcommand{\cH}{{\cal H }}

\newcommand{\ra}{\rightarrow}

\newcommand{\inv}{^{-1}}

\newcommand{\Ker}{\hbox{\rm ker}\,}

\newcommand{\Tr}{{\rm Tr}\,}

\newcommand{\dbar}{{\overline{\partial}}}

\newcommand{\bbar}{{\overline{b}}}
\newcommand{\cbar}{{\overline{c}}}

\newcommand{\qbar}{{\overline{q}}}

\newcommand{\zbar}{{\overline{z}}}

\newcommand{\Lbar}{{\overline{L}}}

\newcommand{\taubar}{{\overline{\tau}}}

\newcommand{\cM}{{\cal M}}
\newcommand{\cF}{{\cal F}}

\renewcommand{\Im}{{\rm Im\,}}

\def\l{\lambda}

\def\t{\tau}
 
\def\F{\Phi}

\def\v{\varphi} 

\def\<{\langle}
\def\>{\rangle}
\newcommand{\hep}[1]{{\tt hep-th/#1}}
\newcommand{\npb}[1]{{Nucl.\ Phys. {\bf B#1}}}
\newcommand{\id}{{\bf 1}}

\def\woinf{W_{1+\infty}}
\def\sint{\int\!}
\def\ll{\left\langle}
\def\rr{\right\rangle}
\def\ddz{d^2\!z\,}
\def\ddw{d^2\!w\,}
\renewcommand{\ll}{\Bigl\langle}
\renewcommand{\rr}{\Bigr\rangle}
\newcommand{\Bbb}{\bf}

\begin{document}
\addtolength{\baselineskip}{.5mm}
\begin{flushright}
August 1996\\
{\sc hep-th/9609022}
\end{flushright}
\vspace{30mm}
\thispagestyle{empty}
\begin{center}
{\Large\sc Chiral Deformations of Conformal Field Theories}
\\[25mm] {Robbert Dijkgraaf}\\[4mm]
{\it Department of Mathematics\\
University of Amsterdam\\ 
Plantage Muidergracht 24\\ 
1018 TV Amsterdam, The Netherlands.}
\\[2cm] {\sc Abstract}
\end{center}

\noindent
We study general perturbations of two-dimensional conformal field
theories by holomorphic fields. It is shown that the genus one
partition function is controlled by a contact term (pre-Lie) algebra
given in terms of the operator product expansion. These models have
applications to vertex operator algebras, two-dimensional QCD,
topological strings, holomorphic anomaly equations and modular
properties of generalized characters of chiral algebras such as the
$W_{1+\infty}$ algebra, that is treated in detail.

\vfill

\newpage

\newsection{Introduction}

In this paper we consider chiral deformation of two-dimensional
conformal field theories.  By a chiral deformation we understand a
field theory with an action of the form
\be
S = S_0 + \int d^2\!z\, A(z),
\ee
where $S_0$ indicates the action of the undeformed conformal field
theory and $A(z)$ is a holomorphic field of arbitrary (integer)
spin. There are various questions and problems posed by this class of
models that we will try to address in this paper. But let us first
indicate some motivations to consider these field theories.

\newsubsection{$(\d\v)^3$ theory and two-dimensional QCD}

We were originally motivated by the following simple two-dimensional
quantum field theory. Consider a real bosonic scalar field
$\v(z,\zbar)$ on a two-dimensional Euclidean space-time with the
topology of a torus and with the following cubic interaction
\be 
S= \int \left(\d \v\dbar\v + \l (\d\v)^3\right).
\label{lagr}
\ee
Here it is important that the interaction term $(\d\v)^3$ is a
holomorphic spin 3 field. This rather uncommon interacting scalar field
theory turns out to be interesting from several points of view.

First, as pointed out by Douglas \cite{douglas},
if we choose the coupling constant $\l=1/N$ this model appears
as an effective string field theory for the two-dimensional QCD string
on a target space torus. Two-dimensional $U(N)$
Yang-Mills theory in the large $N$ limit has been studied in detail by
Gross and Taylor \cite{gross}.  They have shown that the partition
function has a string interpretation in terms of maps of Riemann
surfaces to the target space-time. That is, the partition function has
the characteristic form
\be
Z(\tau,N) = \exp \sum_g N^{2-2g} F_g(\tau),
\label{qcd}
\ee 
where $\tau$ is the (complexified) area of the space-time surface and
the contributions $F_g(\tau)$ `count' the maps of a genus $g$ string
world-sheet to the target space-time. (Roughly,
$F_g(\tau)=\sum_n F_{g,n} q^n$ where $n$ is the degree of the map and
$q=e^{2\pi i \tau}$.)

The general description of the counting functions $F_{g}$ is rather
complicated but can be completely understood in terms of holomorphic
maps \cite{cmr}. (See also \cite{horava}, where a closely related
formalism using harmonic maps is used.) However, 
in the case that the target-space has the
topology of a torus, the combinatorics becomes much more
straightforward and can be summarized by the fact that the string
field theory takes the extremely simple cubic form given above, with
string coupling constant $\l$ given by $1/N$. This remarkable
simplification is very much dependent on the equivalence of the
two-dimensional bosonic scalar field $\v$ with a Dirac spinor
$(b,c)$. In terms of these fermions the $(\d\v)^3$ action simply reads
\be
S= \int \left(b\dbar c + \l\, b \d^2 c\right).
\ee
This is a quadratic action, which account for the solvability of the
model. In fact, similar actions have also appeared in the $c=1$ matrix
model \cite{matrix}. 

This free field theory representation of the QCD string partition
function gives a very simple and elegant formula for the string-loop
genus expansion as a generalized conformal character
\be
Z(\tau,N) = \Tr\left( q^{L_0} e^{H/N}\right),\qquad H=\oint b\d^2 c. 
\ee

It has been noticed that the expansion coefficients $F_g$ that appear
in the perturbation theory in the coupling constant have rather
peculiar modular properties. They are so-called quasi-modular forms
\cite{rudd,texel,zagier}. This raises the interesting issue of what
the modular properties of theories of the $(\d\v)^3$ type are, in
particular how the coupling constant $\l$ transforms. 
This is one of the questions we will answer in generality in this paper.

\newsubsection{Kodaira-Spencer theory}

A second motivation for considering the $(\d\v)^3$ model comes from
topological string theory. In fact, the above model is in many
respects a two-dimensional analogue of the six-dimensional
Kodaira-Spencer field theory that has been introduced as the effective
field theory of a topological string of type B on a Calabi-Yau
three-fold $X$, see \cite{bcov2}. (in this way, the QCD string can
be regarded as a topological string of type A, related by mirror
symmetry of $T^2$.) In the Calabi-Yau case we are dealing
with a six-dimensional Lagrangian, of the form
\be
\int_X \d\v\wedge \dbar\v + \l\, \d\v\wedge \d\v\wedge\d\v,
\ee
where the field $\v$ can be seen as a $(1,1)$ form and the holomorphic
three-form is used to make sense of the cubic interaction. This
quantum field theory is supposed to calculate the instanton sum on the
mirror manifold. Because it has a natural string field theory
interpretation, the obvious problems of this Lagrangian should be
cured using the string regularization. In two-dimensions there is a
unique Calabi-Yau manifold, the torus or elliptic curve. Its mirror
manifold is again an elliptic curve, and the instanton sum is given by
our $(\d\v)^3$ model. See \cite{texel} for more details on mirror
symmetry for elliptic curves in relation to the counting functions of
holomorphic maps.

The two-dimensional model also shares with the Kodaira-Spencer theory
the property that it is superficially non-renormalizable, while finite
in some natural regularization. We will see that in the
two-dimensional model this can be understood in the following way.  At
the expense of introducing contact terms, that we will carefully
analyze, the chiral interactions can be written as contour integrals
of the type
\be
\oint dz\,  (\d\v)^3.
\ee
These contour integrals can be chosen to be non-intersecting, which
eliminates all divergences. In this two-dimensional case this
regularization is much more straightforward than in the
six-dimensional model, where it is supposed the come from string
theory. In fact, one of our motivations was to understand to which
extent the six-dimensional theory has an equally well-defined
perturbation theory. 

The six-dimensional Kodaira-Spencer model is (partially) solvable
through the so-called holomorphic anomaly equation \cite{bcov1}.  We
are therefore also interested in the holomorphic properties of the
above model in terms of the modulus $\t, \taubar$ of the torus. We
will derive an analogue equation for the $\taubar$-derivative of the
partion function, which is a generalization of the holomorphic anomaly
equations derived in six dimensions.

\newsubsection{Chiral algebras and generalized characters}

The $(\d\v)^3$ model can be seen as just a particular example of a
large class of field theories that can be constructed by deforming a
given conformal field theory with an arbitrary chiral operator.
Such an operator has conformal dimensions $(h,0)$ and is therefore not
marginal. Hence the conformal symmetry will be broken. Since the
operator carries spin, the deformation is also not
Lorentz/rotation invariant.

Some examples of perturbations by fields of non-zero spin can be found
in certain models in two-dimensional statistical physics. Here 
rotational symmetry breaking is of course less of a problem. 
A rather famous example is the chiral Potts
model \cite{potts}, that can be considered as a deformation of a 
minimal CFT by an operator of conformal dimensions $({7\over 5},{2\over 5})$ 
and thus of spin one, see \cite{cardy}.

One can also think of these chirally deformed models as field theories
coupled to generalized (constant) background higher spin gauge
fields. This makes the subject of interest in the context of higher
spin analogues of chiral quantum gravity, so-called $W$-gravity
\cite{w-grav}.

The partition functions of such deformed models can be considered as
generalized characters of the chiral algebra underlying the conformal
field theory.  If $R$ is a representation of a vertex operator
algebra $V$ with a basis $H^i$ of commuting Noether charges, then one
can define generalized characters as
\be
\chi_R(\tau,s)={\rm Tr}_{R}\left(q^{L_0-{c\over 24}} e^{s_i H^i}\right).
\ee
with $q=e^{2\pi i \tau}$ and $s_i$ coordinates on the ``Cartan
subalgebra'' of the chiral algebra. These characters carry a
representation of the modular group $PSL(2,\Z)$.  By general
arguments, for a rational conformal field theory, where the
irreducible representations $R_I$ are finite in number, we have a
transformation rule of the form
\be
\chi_I(\tau',s') = \sum_J M_I{}^J \chi_J(\tau,s),
\ee
where $\tau'= (a\tau+b)/(c\tau+ d)$.  One of our aims in this paper
will be to determine how the transformed parameters $s_i'$ are expressed
in terms of the variables $s_i$ and $\tau$ under a modular
transformation. One of our conclusions will be that the variables $s_i$
do not have canonical modular properties, but certain polynomials in
them will transform canonically.

A particular model where all this can be seen in great detail is the
$\woinf$ algebra.  The representation theory of this algebra has been
intensely studied, see \eg  \cite{kac,awata}. In \S4 we will
treat the $c=1$ free field theory realization of $\woinf$.

Finally we mention that instead of looking at holomorphic fields and
characters of chiral algebras, one can also consider $N=2$
superconformal field theories and their elliptic genera
\cite{ell-genus}. These objects behave very much like characters
of holomorphic CFT's. Our results will then apply to the so-called 
``refined elliptic genus'' introduced in \cite{warner}.

\newsection{Chiral Algebras}

We first make a few general comments about chiral algebras of
two-dimensional conformal field theories. For more details about
vertex operator algebras see \eg \cite{axioms,monster}.

\newsubsection{Vertex operator algebras}

Consider a general unitary conformal field theory and let $V$ denote
the space of chiral operators, that is, holomorphic but not
necessarily primary fields $A(z)$ of conformal weight $(h,0)$. This
space $V$ is an infinite-dimensional vector space, naturally graded by
the weight $h
\in \Z_{\geq 0}$ of the operators.  It always contains the identity
$\id$, the unique field of weight zero, and the stress tensor $T$ of
weight two, together with all its descendents.

On this space of chiral operators we have an action of the translation
operator 
\be
\d:={\d\over \d z}=L_{-1}
\ee
that raises the weight of an operator by one. We will consider in
this paper mainly the quotient space
\be
W=V/\d V.
\ee 
One can think of the map $V \ra W$ as
associating to a chiral current $A(z)\in V$ its Noether charge $Q(A)\in
W$, with
\be
Q(A)=\oint {dz\over 2\pi i}A(z).
\label{charge}
\ee
If $A$ has the usual mode expansion $A(z)=\sum_{n\in\Z} A_n z^{-n-h}$,
then this charge is given by $Q(A)=A_{h-1}$. Note that these charges
are the zero modes on the $z$-plane, {\it not} the zero modes $A_0$ on
the cylinder with coordinate $\log z$.

The operator product expansion of two fields $A,B\in V$, denoted here as
\be
A(z)\. B(w) \sim \sum_{n=-\infty}^\infty (z-w)^{-n} (AB)_n(w),
\ee
gives $V$ the structure of a vertex operator algebra. Vertex operator
algebras can be completely axiomatically defined in terms of the
infinite set of operator products $(\.\.)_n$ and the action of the
derivative $\d$ \cite{axioms}. For bosonic fields the operator
products have the symmetry property
\be
(AB)_n = (-1)^n (BA)_n,
\ee
and the conformal weight  of the product $(AB)_n$ is given by $h_A+h_B-n$.

As is well-known, the first order product $(\.\.)_1$ induces a Lie
bracket on the coset space $W$
\be
[A,B]_1 :=(AB)_1\quad ({\rm mod}\ \d).
\label{first}
\ee
We denote this bracket here with a suffix ${}_1$
to stress the fact that it is related to the first order pole in the
operator product. The Jacobi identity only holds up to terms of the
form $\d(\cdots)$, so $W$ and not $V$ is a Lie algebra.  This is the
familiar Lie algebra generated by the corresponding conserved Noether
charges
\be
\Bigl[Q(A),Q(B)\Bigr]=Q\Bigl([A,B]_1\Bigr).
\ee

\newsubsection{Pre-Lie algebra structure}

In the following we will only consider ``abelian'' chiral algebras
where the first order Lie bracket $[\.,\.]_1$ on $W$ is trivial. That
is, we will assume that, possibly after a suitable restriction to a
``Cartan subalgebra,'' the vertex operator algebra $V$ has the
property
\be
(AB)_1 \in \Ker(\d),\qquad \forall A,B \in V.
\label{vanish}
\ee

This assumption has an important consequence, that will play a crucial
role in the rest of the paper. It allows us to define a new product
$W\times W \ra W$, namely
\be
 \nabla_B A := \d\inv(AB)_1
\label{nabla}
\ee
It is straightforward to check that this expression is well-defined 
on equivalence classes, \ie modulo derivatives.
We have written this
product as a covariant derivative, since we will see that it satisfies
all the properties of a flat, torsion-free linear connection for $W$,
if one thinks of $W$ as the space of vector fields on some manifold.

We should make one remark here. As it stands, the definition of the
product $\nabla_B A$ is incomplete. It is only well-defined if central
terms are absent. That is, the identity operator $\id$, with the
complicating property $\d\,\id=0$, should not appear in $\nabla_B A$.
Since the conformal weight of $\nabla_B A$ is given by $h_A + h_B -
2$, this problem only occurs in the case that both $A$ and $B$ are
spin one currents. We therefore restrict ourselves to fields of spin
$h\geq 2$.  We will introduce the spin one fields at a later stage
in \S4.4.

The product $\nabla_B A$ is neither symmetric nor antisymmetric. In
fact, the symmetric part gives the quadratic residue $(AB)_2$ on $W$
\be
(AB)_2=\nabla_A B + \nabla_B A.
\ee
This is the famous commutative but non-associative product that
features in the construction of the Griess algebra --- the fundamental
module for the Monster group \cite{monster}.

The antisymmetric part of $\nabla_AB$ gives rise to a second order Lie
bracket $[\.,\.]_2$ on $W$
\be
[A,B]_2=\nabla_A B - \nabla_B A,
\label{torsion}
\ee
which should be distinguished from the more familiar first order
bracket (\ref{first}).  Equation (\ref{torsion}) can be equivalently
read as saying that the ``connection'' $\nabla$ is torsion-free.
Since we will always assume that the first order Lie bracket vanishes
on $W$, no confusion can arise and we will drop the subscript ${}_2$
from now on. The Jacobi identity for the bracket $[\.,\.]$ follows
again directly from the general Jacobi identity of vertex operator
algebras that gives the relation
\be
[\nabla_A C,\nabla_B C]=\nabla_{[A,B]}C. 
\label{jacobi}
\ee
This can be interpreted as saying that the connection $\nabla$ is
flat, \ie the operator $\nabla_A: W \ra W$ satisfies
\be
[\nabla_A,\nabla_B]=\nabla_{[A,B]}.
\label{flat}
\ee
We stress again that the second order Lie bracket on $W$ is only
well-defined if the first order bracket vanishes.

In terms of mode expansions and Noether charges we simply have
\be
Q(\nabla_A B)=[\nabla_A,Q(B)],\qquad 
\nabla_A =\oint {dz\over 2\pi i}z A(z)=A_{h-2}.
\ee

We also note here that in the special case of the stress tensor $T$
(of spin two) and an arbitrary field $A$ (of spin $h$) we have
\be
\nabla_A T = A,\qquad \nabla_T A=(h-1)A.
\ee

The algebraic structure of a vector space $W$ with a product $\nabla$
satisfying the relations (\ref{torsion}) and (\ref{jacobi}) is
sometimes referred to as a {\it pre-Lie algebra} \cite{gerstenhaber},
since by definition the commutator of the products gives a Lie
bracket. Note that this is a stronger notion than a Lie-algebra: a pre-Lie
algebra is always also a Lie algebra.

Any pre-Lie algebra $W$ acts in two different ways on itself as a Lie
algebra.  First, there is the obvious adjoint action
\be
{\rm ad}_A:\ B \ra [A,B].
\ee
Secondly, there is the fundamental action
\be
\nabla_A:\  B \ra \nabla_A B.
\label{fun}
\ee
A simple (and canonical) example of a pre-Lie algebra is the
Lie algebra of vector fields on $\R^n$ with $\nabla$ the trivial
connection (or more general a manifold with a flat metric).
If we trivialize a vector field $A$ on $\R^n$ in flat coordinates
$x^i$ as $A=A^i \d_i$, then we can treat the components as
functions:
\be
(\nabla_A B)^j= A^i\d_iB^j.
\ee

\newsection{Chiral Conformal Perturbation Theory}

We now turn to the more general problem we want to address in this paper,
the discussion of chiral deformations of conformal field theories. 

\newsubsection{Modular invariance}

Consider the partition function $Z(\t,\taubar)$ of a general unitary
conformal field theory with (abelian) chiral algebra $V$ on a torus or
elliptic curve $E$, with modulus $\t\in\H$ in the upper-half plane. We
want to deform the action $S_0$ of the conformal field theory by
adding to it a term of the form
\be
\int_E \ddz A(z)
\ee
with $A(z)\in V$ a chiral current. Since total derivatives integrate
to zero on a compact space --- a property we will carefully preserve
in the regularization procedure --- we can consider $A$ to be actually
an equivalence class in the quotient space $W=V/\d V$ of fields modulo
total derivatives. We will refer to $W$ as the Cartan algebra.

If $\F_i$ is basis for $W$, the general form of the
chiral perturbation of the action will take the form
\be
S=S_0 -\int {d^2\!z\over 2\pi\t_2} t^i \F_i(z).
\label{action}
\ee
We can think of the $t^i$ as constant background gauge fields.  The
term $\t_2=\Im\t$ is added to ensure proper modular weights for the
coupling constants $t^i$. In fact, under a modular transformation, that
acts on the modulus $\tau$ by fractional linear transformation
\be
\t \ra \t'={a\t + b\over c\t +d},\qquad \twomatrixd abcd \in PSL(2,\Z),
\label{modtrans}
\ee
we have the following transformation rule of the linear coordinate
$z\in E$:
\be
z \ra z'={z\over c\t+d}.
\ee
Consequently a chiral field $\F_i(z)(dz)^{h_i}$ of {\it conformal}
weight $h_i$ transforms as
\be
\F_i(z) \ra (c\t+d)^{h_i} \F_i(z')
\ee
and so also has {\it modular} weight $h_i$. Since the action
(\ref{action}) should be modular invariant, the coupling constant
$t^i$ dual to $\F_i$ transforms as a modular forms of weight $-h_i$,
\be
t^i \ra (c\t+d)^{-h_i} t^i.
\ee

In more fancy terms: the family of perturbed field theories
parametrized by the variables $t^i$ forms a non-trivial vector bundle
over the genus one moduli space $\cM_1=\H/PSL(2,\Z)$.

\newsubsection{Contact terms}

In order to make rigorous sense of the deformed model in terms of
perturbation theory around the original undeformed
conformal field theory, one has to make sense out the following 
generating functional of correlation functions
\be
Z[t]=\ll \exp \int {\ddz\over 2\pi\tau_2} t^i \F_i(z)\rr.
\ee
When the exponential is expanded, we encounter terms of the form
\be
\ll \cdots \int {\ddz\over 2\pi\tau_2} A(z) \int {\ddw\over2 \pi\tau_2 }
B(w)\cdots\rr
\ee
for some fields $A(z),$ $B(w)$ and there will be singularities in the
integrand for coinciding position variables $z=w$. We will have to
prescribe how to integrate over these poles. 

Our principal value prescription will be the one proposed by Douglas
in ref.\ \cite{douglas}. In his proposal one writes
\be
A(z)=\dbar C,\qquad C(z,\zbar)=(\zbar -z)A(z),
\ee
and subsequently applies Stokes' theorem to the integrals, 
where little disks are cut out around the positions of
the operator insertions.
Since the operator $C$ is not single-valued, we pick up both a period
contribution from the multi-valuedness and a residue contribution
from the poles at the punctures. These contributions
can however be explicitely evaluated in
terms of the operator product coefficients, with the result
\be
\int  {\ddz\over 2 \pi\tau_2} A(z)\. B(w)=
\int_0^1 {dz \over 2\pi} A(z) \. B(w) + {1\over 2\t_2}\left(w\. 
(AB)_1(w)+(AB)_2(w)\right).
\label{contact}
\ee
The last two contributions on the right-hand side can be interpreted
as contact terms due to first and second order poles in the operator
product respectively.  Since in our formulas all operators are
integrated in the end, we can consistently work modulo $\d$ and freely
perform a partial integration in the variable $w$. That allows
us to replace the second and third term on the right-hand side of
(\ref{contact}) by the term
\be
c(A,B)={1\over 2\t_2}\left[(AB)_2 - \d\inv(AB)_1\right]={1\over2
\t_2}\nabla_A B.
\label{alg}
\ee
Here we used definition (\ref{nabla}) of the pre-Lie structure of the
Cartan algebra $W$. According to the discussion in \S2 
the contact term (\ref{alg}) is well-defined modulo total
derivatives, \ie makes sense on the quotient space $W=V/\d V$. It
has no obvious symmetry properties under interchange of
the two arguments. 

More precisely, if we introduce the following
short-hand notation for surface and contour integrals respectively
\be
\sint A = \int  {\ddz\over 2 \pi\tau_2} A(z),\qquad
\oint A = \int_0^1 {dz \over 2\pi} A(z),
\ee
we find that the following relation is valid within correlation
functions
\be
\sint A \sint B = \oint A \sint B + \sint c(A,B) + \ldots
\label{rec}
\ee
Here $c(A,B)$ indicates the contact term between the fields $A$ and
$B$ and the ellipses represent similar terms if extra fields like $B$
are present. We notice that all contact terms disappear in the limit
$\t_2\ra\infty$.

We can use equation (\ref{rec}) to recursively eliminate all surface
integrals in terms of contour integrals. It is instructive to work
this out explicitly for correlators with a small number of operators.
Using the identities of the previous section one finds that everything
can be expressed in terms of the second order operator product
$(\.\.)_2$. For example, the two-point function satisfies
\be
\ll \sint A \sint B\, \rr  = 
\ll \oint A \oint B \,\rr  + {1\over2\t_2}
\ll \oint (AB)_2\,\rr ,
\ee
and similarly the three-point function satisfies
\ba
\ll   \sint A \sint B \sint C\,\rr  \is  
\ll   \oint A \oint B \oint C\,\rr \nonu
& & +  {1\over2\t_2} \ll  \oint A \oint (BC)_2 + \mbox{\it cyclic}\,\rr \nonu
& & \qquad +  {1\over8\t_2^2} \ll  \oint (A(BC)_2)_2 + \mbox{\it cyclic}\,\rr .
\ea

\newsubsection{Reparametrization of coupling constants}

We can now reformulate our deformation problem as follows. The most
general action with chiral interactions we want to consider is of the
type
\be
S=S_0 -  \sint t^i\F_i.
\ee
We have seen in the previous subsection that these chiral interactions
can be rewritten in terms of contour integrals at the expense of
introducing contact terms. In that way the deformed Lagrangian $S$ is
rewritten in terms of a deformed Hamiltonian $H$, where we integrate
chiral fields over a space-like contour,
\be
H = H_0 + \oint  s^i\F_i.
\ee
The Hamiltonians $\oint \F_i$ are the conserved charges or zero modes of the
chiral currents $\F_i$ on the cylinder. They should not be confused with the 
charges introduced in (\ref{charge}). If a chiral field $A$ of spin
$h$ has a mode expansion $A(w)=\sum_n A_n w^{-n-h}$ on the plane, with
coordinate $w=e^{2\pi i z}$, then the corresponding Hamiltonian is
given by\footnote{Because of the conformal anomaly $c$,  this becomes 
for the stress-tensor $\oint T=-2\pi L_0 +{\pi c\over 12}$.}
\be
\oint {dz\over 2\pi} A = i(2\pi i)^{h-1} A_0.
\ee
By the familiar contour deformation argument, these Hamiltonians commute,
\be
\Bigl[\oint \F_i,\oint \F_j\Bigr]=0,
\ee
because of our assumption (\ref{vanish}) that the chiral algebra is
abelian. There is consequently no ambiguity in the choice of contours.
They can be chosen to be non-intersecting in some arbitrary time
ordering. Therefore, in Hamiltonian perturbation theory the model is
perfectly well-defined. The partition function can be computed in the
operator formalism as a trace in the Hilbert space $\cH$ of the
conformal field theory,
\ba
Z[\t,\taubar;s] \is \ll \exp  \oint s^i \F_i\rr \nonu
\is \Tr_{\strut\cH}\Bigl[q^{L_0-c/24}\qbar^{\Lbar_0-c/24} 
\exp \oint s^i\F_i\Bigr],
\ea
with $q=e^{2\pi i \t}.$ 
The partition function $Z[\tau,\taubar;s]$ should be considered in the limit 
$\taubar \ra -i\infty$ as a generalized character of the chiral
algebra $V$.

Note that the interactions that we have added have weights greater
than $2$ and are strictly speaking nonrenormalizable. However, because
of holomorphicity, they are effectively integrated only over
one-dimensional cycles. Therefore, they do not give rise to the
expected divergences of nonrenormalizable interactions.

In our notation we already anticipated that the parameters $t^i$ in
the Lagrangian and the parameters $s^i$ in the Hamiltonian will
differ. Indeed, due to the effect of the contact terms, the coupling
constants $s^i$ will in general be some non-trivial function of the
so-called canonical coordinates $t^i$ that appear in the
Lagrangian \cite{bcov2}
\be
s^i=s^i[t^j].
\label{st}
\ee
The fact that contact terms induce a reparametrization of the space of
coupling constants is a familiar phenomenon in conformal field theory 
\cite{kutasov}.  It is precisely in the appearance of contact terms
that the superficial non-renormalizability of the model (re)emerges.
Indeed, from (\ref{alg}) we see that the weight of the contact term
$c(A,B)$ is $h_A+ h_B -2$. Since we have to add this higher spin field
to the Lagrangian with a non-zero coupling constant, this gives a
cascade of terms of higher and higher dimension. For example, two spin
3 fields can produce a spin 4 field in their contact term, which on
its term can produce a spin 5 field, {\it etc. etc.}

Another way to see the necessity of a reparametrization of the
coupling constants, is that the variables $s^i$ have {\it a priori} no
obvious modular properties. We explicitly have broken the modular
invariance of the model by picking a preferred cycle (in this case the
$a$-cycle, $z$ real, a constant time-slice) on the torus.
Equivalently, in the Hamiltonian formalism, by a choice of a
time-direction, we break global diffeomorphism invariance.

\newsubsection{A simple example: the stress-tensor deformation}

This reparametrization effect is perhaps most familiar in the case of
a perturbation by the stress-tensor, which is simply a deformation of
the metric,
\be
\delta S = - t  \sint T(z).
\label{zz}
\ee
This can be related to a shift in the Hamiltonian
\be
\delta H= s \oint  T(z)= - 2\pi s L_0
\ee
for a particular function $s(t)$ as follows. 
(See also \cite{bcov2} where this example is discussed in the general
context of deforming the complex structure of a Calabi-Yau manifold.)
The deformation (\ref{zz}) translates into a deformation
of the $\dbar$-operator of the form
\be
\dbar_\mu=\dbar+\mu\,\d 
\ee
with Beltrami differential $\mu= -t/2\t_2$. If we write the complex
coordinates $z=x_1+\t x_2$, $\zbar=x_1+\taubar x_2$, so that
\be
\d = {\taubar \d_1 - \d_2 \over \taubar-\t},\qquad 
\dbar = {-\t\d_1 + \d_2 \over \taubar -\t},
\ee 
we see that this corresponds to a deformation of the modulus $\t_\mu$
given by
\be
\t_\mu= \t +{2i\t_2 \mu \over 1 - \mu}.
\ee
(In these formulas the complex-conjugate is left unchanged,
$\taubar_\mu=\taubar$.)  Since the variable $s$ is given by
$s=-i(\t_\mu-\t)$, we find $s={2\t_2 \mu}/( 1 - \mu).$ Therefore the
coupling constants $s$ and $t$ are related as
\be
{s\over 2\t_2}+1=\left(1-{t \over 2\t_2}\right)\inv.
\label{s-t}
\ee
We will have a chance to verify this relation in a moment. The
variable $t$ (or $\mu$) is the so-called canonical coordinate,
centered at $\t$, and the variable $s$ (or $\t_\mu$) can be thought of
as the canonical coordinate centered at $\t=i\infty$ \cite{bcov2}.  We
will generalize this point of view to arbitrary coupling constants.
In the limit $\t_2 \ra \infty$ all contact terms disappear, and we
find that in a perturbation around that point the coupling constants
are simply identical, $s^i=t^i$, so $s^i$ is indeed the canonical
coordinate at infinity.

\newsubsection{Differential equations and recursion relations}

We now wish to calculate the relation between the two sets of coupling
constants $s^i$ and $t^i$ that follows from the equality of the
partition sums
\be
\ll \exp\sint  t^i \F_i\rr=\ll\exp \oint  s^i\F_i\rr
\ee
The relation between the coupling constants can in principle
be solved by considering the more general generating function of two
sets of variables
\be
Z[s,t]=\ll\exp \left(\oint s^i \F_i+ \sint t^i\F_i\right)\rr.
\ee
We want to find the relation between $s^i$ and $t^i$ such that
\be
Z[s,0]=Z[0,t].
\ee
We will demonstrate that, as 
a consequence of the contact term relation (\ref{rec}), the
generalized partition function $Z[s,t]$ satisfies a set of
differential equations, that allows us to solve the dependencies.

To fix notation, we will write the connection
$\nabla$ in terms of our basis $\F_i$ as
\be
\nabla_i\F_j= c_{ij}{}^k\F_k.
\ee
Consider now the linear first-order differential operators
$L_i^{(s)},L_i^{(t)}$ with
\be
L^{(s)}_i = \sum_{j,k}  c_{ij}{}^k s^j \pp{s^k},\qquad L_i^{(t)} =
\sum_{j,k} c_{ij}{}^k  t^j\pp{t^k}.
\ee
They form a representation of the Lie algebra $W$, as introduced in \S2,
\be
[L_i,L_j]= f_{ij}{}^k L_k,\qquad f_{ij}{}^k=c_{ij}{}^k-c_{ji}{}^k.
\ee
We now claim that the partition function $Z[s,t]$ satisfies the
following linear differential equation, that we will call the ``master
equation''
\be
\Pp Z {t^i}=\left[\pp{s^i} + {1\over 2\t_2}
\left(L_i^{(s)}+L_i^{(t)}\right)\right]Z.
\label{z}
\ee

Let us first try to explain the derivation of this relation in words.
Differentiating the generating function $Z$ with respect to $t^i$
``brings down'' the surface integral $\sint \F_i$. According to the
fundamental contact term relation (\ref{rec}) we can write this as a
contour integral $\oint \F_i$, \ie a differentiation with respect to
$s^i$, plus additional terms coming from first and second order poles
in the operator product of the field $\F_i$ with the various other
fields.  This contact term has the form (\ref{alg}) and is valid {\it
both} for the surface and contour integrals. More explicitly, in
terms of a particular term in the generating function $Z$,
\smallskip

\ba
\ll \sint \F_i \prod_{m\in M} \sint \F_m \prod_{n\in N} \oint \F_n\rr \is 
\ll \oint \F_i  \prod_{m\in M} \sint \F_m \prod_{n\in N} \oint \F_n\rr 
\nonumber \\[4mm]
& & \hspace{-10mm} + \sum_{j\in M} {1\over 2\t_2} \ll 
c_{ij}{}^k \sint
\F_k \prod_{m\in M-j} \sint \F_m \prod_{n\in N} \oint \F_n\rr 
\label{recursion}
\\[4mm]
& & + \sum_{j\in N}{1\over 2\t_2} \ll c_{ij}{}^k \oint \F_k
\prod_{m\in M} \sint \F_m \prod_{n\in N-j} \oint
\F_n\rr.\qquad\nonumber
\ea
\smallskip

\noindent 
Here $M,N$ are two subsets of indices. By an argument familiar from
the theory of two-dimensional topological gravity \cite{topgrav} the
effect of this contact term algebra is represented on the generating
function $Z$ by the differential operators $L_i^{(s),(t)}$.

The master equation (\ref{z}) can be used to eliminate the variables
$t^i$ in favor of $s^i$ once the structure of the pre-Lie algebra $W$
is given.  We will illustrate this with a concrete model in the next
section, but as a warming-up, let us first reconsider the perturbation
with the stress-tensor $T$ with two couplings $t$ and $s$ discussed in
\S3.2.  Since we have the simple relation $\nabla_TT=T$, in this case
the master equation reads
\be
\pp t Z =\left(\pp s + {1\over 2\t_2} s \pp s +{1\over 2\t_2} 
t \pp t\right)Z.
\ee
If we introduce new variables $a=1+{s\over 2\t_2},$ $b= 1-{t \over 2\t_2}$,
the master equation  reduces simply to
\be
\left(a\pp a + b\pp b\right) Z=0,
\ee
so that $Z[a,b]=Z[a/b].$ Together with the appropriate initial
conditions this tells us that $Z[a,0]=Z[0,b]$, where the couplings
$a,b$ are related via $a=1/b$ or
\be
{s\over 2\t_2}+1=\left(1-{t \over 2\t_2}\right)\inv.
\ee
This is indeed the relation we found in (\ref{s-t}).

As this example shows clearly, at this point it is advantageous to
introduce a differently normalized
set of coupling constants $a^i,$ $b^i$ defined by
\be 
a^i = {s^i \over 2\t_2} + \delta^{i,T},\qquad
b^i = -{t^i\over 2\t_2} + \delta^{i,T}.
\label{redef}
\ee
Here $i=T$ labels the coupling to the stress-tensor. If we make use of
the identity
\be
\nabla_AT=A,\qquad \forall A\in W, 
\ee
one can verify that after this shift the constant terms in (\ref{z})
disappear and in terms of the new coupling constants $a^i,$ $b^i$ we
have the simplified equation
\be
\left(L_i^{(a)}+ L_i^{(b)}\right)Z[a,b]=0,
\label{a-b}
\ee
with an expansion around $a^k=b^k=\delta^{k,T}$. 

Written like this, the master equation has a simple
interpretation. Remember that the pre-Lie algebra $W$ carries, besides
the adjoint representation, a second, fundamental representation
(\ref{fun}) of the underlying Lie algebra. With this representation,
the above equation simply states that the partition function
$Z:W\times W \ra \C$ is an invariant function.

\newsection{The $c=1$ Model}

We will now turn to the example that motivated the above discussion:
the $c=1$ bosonic field $\v$ with interactions of the form $(\d\v)^n$.
Here it is most convenient to use the equivalent fermionic
formulation in terms of a spin $\half$ $(b,c)$ system or Dirac fermion.
(For the partition function we
should remember to integrate in the end over the spin
structures if we wish to obtain the bosonic partition function.)

\newsubsection{$\woinf$ algebra}

The free boson or Dirac fermion forms a $c=1$ representation of the
$\woinf$ vertex operator algebra that is generated as an algebra by
the local chiral fields
\cite{awata}
\be
\F^{p,q}(z)=\d^pb\,\d^qc
\label{pq}
\ee
of spin $h=p+q+1$. For a given spin $h\geq 1$, only one particular
(rather complicated) linear combination of these operators is actually
a primary field. However, since we will only be interested in the
algebra $W$ of operators modulo total derivatives, we can represent
the unique primary field $\F^n$ of weight $h=n+1$ by the class of
operators\footnote{In order to eliminate possible confusing notation,
we will raise/lower indices in this section compared with the previous
section.}
\be
\F^n(z) = - b\d^nc = (-1)^{n-1} \d^nb c\quad ({\rm mod}\ \d).
\ee
In terms of the bosonic field $\v$ this field is represented by
\be
\F^n(z)={1\over n+1}(-i\d\v)^{n+1}\quad ({\rm mod}\ \d).
\ee
For the moment we do not want to consider the $U(1)$ current
$\F^0=-bc$ and therefore only study deformations using the
$W_\infty$-piece, generated by the currents $\F^n$ with $n \geq 1$ of
spin $h=n+1\geq 2$.

In this particular example the connection $\nabla$ that determines the
contact terms is easily computed using the operator product expansion of
the fermion bilinears, with the result
\be
\nabla^m\F^n = n \F^{m+n-1}.
\ee
{}From this it follows that the second order operator product and Lie bracket
are given by
\be
(\F^m\F^n)_2 = (m+n)\F^{m+n-1},\qquad
[\F^m,\F^n] = (n-m)\F^{m+n-1}.
\ee
So the underlying Lie algebra is the (positive part of the) Virasoro algebra
(actually the Witt algebra) and the pre-Lie algebra $W$ is isomorphic
to the space of holomorphic vector fields on the complex plane
vanishing at the origin,
\be
\F^n \sim x^n {\d\over \d x},\qquad n\geq 1,
\ee
with the trivial connection. This representation of $W$ will be useful
in the following.

\newsubsection{The deformed model}

Let us now turn to the family of perturbed conformal field theories.
In terms of the fermions the most general action we want to consider
is of the form
\be
S={1\over \pi}\int \d^2z \left(b\dbar_tc + \bbar \d \cbar\right),
\ee
with $\dbar_t$ the deformed ``$\dbar$-operator'' parametrized by the
coupling constants $t_n$, $n\geq 1$ (or $b_n$ as in (\ref{redef})) as
\ba
\dbar_t \is \dbar-\sum_{n=1}^\infty {t_n\over 2\t_2}\d^n \nonu
\is \dbar -\d + \sum_{n=1}^\infty b_n\d^n.
\ea
Equivalently, in terms of the bosonic field $\v$, we have an action
with general $(\d\v)^n$ interactions
\be
S={1\over \pi}\int d^2z \left(\half \d\v\dbar\v - V(-i\d\v)\right),
\ee
with potential\footnote{Interactions of this type have also appeared in
\cite{eguchi} in the context of the $c=1$ string. 
Here the following `duality' was pointed out: Let $H_n = \oint
(i\d\v)^{n+1}$ and consider the map $w=i\d\v(z)$. Then its (formal)
inverse $z=i\d\chi(w)$ has a mode expansion $\chi(w) =\sum_n H_n
w^{-n}$. So the interchange of `base' and `target' manifold,
interchanges the zero-modes of spin $n$ fields with the $n$-th order
modes of a spin zero field.}
\ba
V(x) \is \sum_{n=1}^\infty {t_n\over 2\t_2}{x^{n+1}\over n+1} \nonu
\is  x- \sum_{n=1}^\infty b_n {x^{n+1}\over n+1}.
\ea
In this case the differential operators $L_n^{(a)},L_n^{(b)}$
$(n=0,1,\ldots)$ are given by Virasoro generators\footnote{Here, by a
slight misuse of notation, the Virasoro generator $L_n$ corresponds to
the field $\F^{n+1}$ of conformal dimension $n+2$.}
\be
L_n^{(a)} = \sum_{k=1}^\infty k a_k\pp{a_{k+n}},\qquad L_n^{(b)} =
\sum_{k=1}^\infty k b_k\pp{b_{k+n}},
\ee
with Virasoro algebra
\be
[L_n,L_m]=(n-m)L_{n+m}.
\ee
The master equation (\ref{a-b}) now reads
\be
\left(L_k^{(a)} + L_k^{(b)}\right)Z=0,\qquad k=0,1,\dots,
\label{vir}
\ee
where we should remember to expand around $a_k=b_k=\delta_{k,1}$.

\newsubsection{Solution of the master equation}

The master equation can be solved as follows. Introduce the holomorphic
functions $a(x),b(x)$, vanishing at $x=0$, with Taylor expansions
\be
a(x)=\sum_{n=1}^\infty a_n x^n,\qquad
b(x)=\sum_{n=1}^\infty b_n x^n.
\label{function}
\ee
On these functions the Virasoro generators $L_k$ act of course as the
vector fields
\be
L_k =x^{k+1}\pp x.
\ee
These vector fields generate the holomorphic diffeomorfisms $f:\C\ra\C$
of a neighbourhood of $0$ that leave the origin fixed.  Condition
(\ref{vir}) now expresses the fact that the {\it functional} $Z[a,b]$
is invariant under these diffeomorfisms
\be
Z[a\circ f,b\circ f ]=Z[a,b].
\ee
This fact can be used to determine the function $a$ in terms of the
function $b$ directly.
Let $\id$ be the identity map, $\id(x)=x$. Now choose $a=\id$ and
$f=b\inv$, the inverse map. (This inverse always exists as a 
power series expression.) Then the above equation implies
\be
Z[\id,b]=Z[b\inv,\id].
\label{invers}
\ee
So the functions $a$ and $b$ are simply each other's inverses,
\be
a(b(x))= x.
\ee
This relation can be easily expanded in terms of Taylor coefficients
\be
a_1 = {1\over b_1},\quad
a_2 = -{b_2\over b_1^3},\quad
a_3 = -{b_3 \over b_1^4} + {2b_2^2\over b_1^5},\quad etc.
\ee
The first relation has been established for a general CFT in
(\ref{s-t}). Note now that in terms of the coupling constants
$s_n,t_n$ we have the relation
\be
y=x + \sum_n {s_n\over 2\tau_2} x^n,\qquad
x=y - \sum_n {t_n\over 2\tau_2} y^n.
\label{rel}
\ee
In terms of the coefficients this gives 
\be
s_1 = {t_1 \over 1 - {t_1\over 2\t_2}},\quad
s_2 =  {t_2 \over ( 1 - {t_1\over 2\t_2})^3},\quad
s_3 = {t_3 \over ( 1 - {t_1\over 2\t_2})^4} +
 {t_2^2 \over \t_2 ( 1 - {t_1\over 2\t_2})^5}
,\quad etc.
\ee
These relations have of course also a straightforward interpretation
in terms of tree level Feynman diagrams, with $n$-th order vertices
labeled by $t_{n-1}$.

\newsubsection{Spin one fields}

Until now we have only considered deformations by fields of spin two or
greater. One of the reasons for this was that definition (\ref{nabla})
of the contact term product $\nabla_B A$ was ill-defined in case both $A$
and $B$ have spin one. However, it is not very difficult to include
the spin one fields too, as we will now illustrate for the $c=1$
model.

Consider the current
\be
\F^0=-bc=-i\d\v,
\ee
with corresponding coupling constants $s_0,t_0$. No problems
arise with the previous formalism if we consider contact terms of
$\F^0$ with fields $\F^n$ with $n>0$. We simply have
\be
\nabla^0\F^n=n\F^{n-1},\quad
\nabla^n\F^0=0,\qquad \ n\geq 1.
\ee
so that in particular
\be
[\F^0,\F^n]=n.
\ee
This implies that (taking into account the
shift by one that we use in our notation) we have to add the extra
generator
\be
L_{-1} = \sum_{k=1}^\infty k t_k\pp{t_{k-1}}
\ee
to our Virasoro algebra. Note that still no central charge term in
the Virasoro algebra appears.

The mutual contact terms of the spin one fields are a bit more
subtle. They are of course given in terms of the second order operator
product
\be
\F^0(z)\F^0(w)\sim {1 \over (z-w)^2}.
\ee
However, in our formalism we now have to distinguish between the
situation where this second order pole is integrated over a contour or
over the surface of the torus. Only in the latter case do we get a
contribution in the recursion relation (\ref{recursion}).

All of this combinatorics can be collected in the following addition
to our master equation:
\be
\Pp Z {t_0}=\left[\pp{s_0} + {1\over 2\t_2}\left(L_{-1}^{(s)}+
L_{-1}^{(t)}\right) + {t_0\over 4\pi\tau_2} \right]Z.
\label{spinone}
\ee
After the usual shift (\ref{redef}), where we replace the variables
$s_i,t_i$ by the variables $a_i,b_i$, the above relation reduces to
the extra constraint
\be
\left[L_{-1}^{(a)} + L_{-1}^{(b)} - {\t_2\over 2\pi}b_0\right] Z=0.
\ee

It is not difficult to derive the solution of this condition by
a similar argument as in the previous subsection.
Extending the definition of the functions $a(x),b(x)$ in
(\ref{function})  by including the
constant terms $a_0,b_0$, we find after some algebra that relation
(\ref{invers}) is generalized to
\be
Z[\id,b]= \exp\left( {\t_2\over 2\pi}B\right)Z[b\inv,\id]
\ee
with constant
\be
B = \int_0^{b\inv(0)}\! b(x)dx=\int_0^{b(0)} b\inv(y)dy.
\ee

It might be instructive to consider this relation in
the case that only the spin one fields are included.
In that case $t_0=s_0$. Since $b(x)=x - {t_0\over 2\tau_2}$ and 
$a(x)=b\inv(x)=x+{t_0\over 2\tau_2}$, the constant $B$
is given by $B={t_0^2/4\tau_2^2}$.
The master equation now reduces to the statement
\be
\ll \exp{\int t_0\F^0} \rr = \exp\left(\tau_2 t_0^2\over 8\pi
\tau_2\right) \ll \exp {\oint t_0\F^0} \rr
\label{uone}
\ee

This equation is familiar for the bosonic model. Then the current 
$\F^0=-i\d\v$ is a total derivative and integrates to zero so that
$\int \F^0 =0$. On the other hand the $U(1)$ zero-mode
$J_0=-i\oint \F^0$ is non-vanishing and 
can be inserted in the partition function
(with $t_0=2\pi i z$)
\be
Z(z)=\Tr\left(e^{2\pi i z J_0} q^{L_0-{1\over 24}}
q^{\Lbar_0 - {1\over 24}}\right).
\ee
It gives the well-known result
\be
Z(z)=e^{-{\pi z^2\over 2\tau_2}}\, Z(0),
\ee
which is in accordance with equation (\ref{uone}).

\newsubsection{Modular properties of $\woinf$ characters}

Let us now discuss the implications of all this for the transformation
rules of the $\woinf$ characters. We define the conserved charges
\be
H^n = \oint \Phi^n,
\ee
and consider the generalized character (with 
spin structure $\alpha,\beta=0,\half$)
\be
\chi(\tau,s)= {\rm Tr}_{\strut\cF}
\left(y^{H_0} q^{L_0- {1\over 24}} e^{s_n H^n}\right).
\ee
Here we also added the $U(1)$ charge $H^0=iJ_0$ with $y=e^{2\pi
\beta}$.  The trace is taken in the free fermion Fock space $\cF$ with
boundary conditions $b(e^{2\pi i}z)=e^{2\pi i\alpha} b(z)$, $c(e^{2\pi
i}z)=e^{-2\pi i\alpha} c(z)$. As is well-known the three even spin
structures on the torus will transform into each other, while the odd
one is invariant. These transformations of the spin structures are
always implicitly understood in the following. Alternatively, one can
also restrict the modular transformation to the subgroup $\Gamma^0(2)$
which leaves the spin structure invariant.

Of course, this $\woinf$ character can be easily evaluated, since the
Hamiltonians act diagonal in the fermion basis. We find \cite{awata}
\be
\chi(\tau,s)=q^{-{1\over 24}} 
\prod_{p\in \Z_{\geq 0}+\alpha}
\left(1 + y q^p e^{iS(2\pi ip)}\right)
\left(1 + y\inv q^p e^{-iS(-2\pi i p)}\right),
\ee
with the notation
\be
S(p) = \sum_{n\geq 0} s_n p^n.
\ee
We are interested in the modular properties of this character. Under a
modular transformation
\be
\tau \ra \tau'={a\tau +b\over c\tau + d}
\ee
we will have
\be
\chi(\tau,s) \ra \chi(\tau',s'),
\ee
where we want to determine the transformation rule of the transformed
variables  $s'_n$. 

As we hope has become clear in \S3, our philosophy is that the
Hamiltonian variables $s_n$ do not have canonical transformation
properties, but the Lagrangian variables $t_n$ in contrast do
transform simply, {\it viz.}\ with modular weight $-(n+1)$
\be
t_n \ra t'_n={t_n \over (c\tau+d)^{n+1}}.
\ee
The transformation properties of the coefficients $s_n$ can now be
read off from the relations (\ref{rel}). Unfortunately, we have not
found an elegant closed expression (although integral formulas are
easily written down) for $s'_n$. But for the first few terms we find
(with $s_i,t_i=0$ for $i=0,1$) that $s_2$ still has modular weight
$-3$ but that $s_3$ has a more complicated transformation behaviour
\ba
s_2 & \ra & {s_2 \over (c\tau+d)^3},\nonu
s_3 & \ra & {s_3 \over (c\tau+d)^4} - {2ic s_2^2 \over (c\tau+d)^5}.
\ea
This implies that the expansion coefficients of the character
\be
\chi(\tau,s) = \sum \chi^{n_1,\ldots,n_k}(\tau) 
s_{n_1} \cdots s_{n_k}
\ee
have corresponding modular properties. In fact, by generalizing
the arguments of \cite{zagier} one can prove that the coefficients
$\chi^{n_1,\ldots,n_k}(\tau)$ will transform as quasi-modular forms,
of weight $\sum_i (n_i+1)$.

\newsubsection{The holomorphic anomaly equation}

As we mentioned in the \S1, it is of interest to consider the
holomorphic anomaly equation of \cite{bcov1,bcov2} in this
context. That is, we consider the partition function
$Z[\tau,\taubar;s]$ of the perturbed model and try to derive an
equation for the anti-holomorphic derivative $\d Z/\d \taubar$. This
is most easily done in terms of perturbation theory of the bosonic
model.

As we will explain in a moment, for our purposes it is most convenient
to work in terms of the Hamiltonian variables $s_n$.
So starting point is the action
\be
S=\int{d^2z\over 2\pi} \d\v\dbar\v + 
\oint \sum_n {s_n \over n+1}(\d\v)^{n+1}.
\ee
The couplings $s_n$ will be treated perturbatively. Since the
interaction terms are chiral, we will only use
the {\it holomorphic} propagator given by
\ba
P(z)= \ll\d\v(z)\d\v(0)\rr \is
 - \wp(z) + {\pi^2\over 3} E^*_2\nonu
\is \d_z^2 \log \theta_1(z) + {\pi\over \tau_2}.
\ea
Here $\wp(z)$ is the Weierstrass function
\be
\wp(z) = {1\over z^2} + \sum_{(m,n)\neq (0,0)}
\left( {1\over (z-(m\tau+n))^2} - {1\over (m\tau+n)^2}\right),
\ee
and $E_2^*$ is defined as
\be
E^*_2(\tau,\taubar) = E_2(\tau) - {3\over \pi \tau_2}
\label{Etwo}
\ee
with the Eisenstein series
\be
E_2(\tau)= 1 - 24 \sum_{n\geq 1} {n q^n \over 1 - q^n}.
\ee
Because of the zero mode contribution, the propagator has an explicit
$\taubar$-dependence
\be
{\d\over \d\taubar} P(z)={i\pi \over 2 \tau^2_2}.
\label{prop}
\ee

The interaction vertices are explicitly holomorphic in $\tau$,
since we have chosen to write them in terms of contour
integrals.  So the only non-holomorphic dependency of the perturbative
expansion of the partition function arises from the propagator.
Because of the simple relation (\ref{prop}), there is a graphical
representation of the action of $\taubar$-derivative on the Feynman
graphs: it simply removes an edge. In fact, here we have to
distinguish two cases. If the propagator connects two distinct
vertices of order $k+1$ and $l+1$ with coupling constants $s_k$ and
$s_l$, these vertices will be replaced by vertices of order $k$ and
$l$ respectively. Taking into account that in the canonical $\woinf$
normalization a vertex of order $k+1$ is weighted by a factor
$1/(k+1)$ instead of the usual symmetry factor $1/((k+1)!$, this
action is represented on the partition function by the differential
operator
\be
\sum_{k,l\geq 1} ks_k \, ls_l \, {\d^2 \over \d s_{k-1} \d s_{l-1}}.
\ee
Similarly, if the propagator begins and ends at a vertex of order
$k+1$, the $\d/\d\taubar$ will remove two outgoing edges
and reduce this vertex to order $k-1$.
The corresponding differential operator is
\be
\sum_{k\geq 2} k(k-1) s_k {\d\over \d s_{k-2}}
\ee
Combining everything and taking into account  the correct constant
of proportionality, we conclude that the anti-holomorphic dependence
can be summarized in the simple equation for the partition function
$Z[\tau,\taubar;s]$, considered as a generating function of Feynman
graphs:
\be
{\d Z \over \d\taubar}={i\pi \over 2\t_2^2} 
\left(L_{-1}^{(s)}\right)^2 Z.
\label{hol-ano}
\ee
Here $L_{-1}$ is the Virasoro generator that we introduced in \S4.4
\be
L_{-1}^{(s)}=\sum_{n\geq 0} n s_n \pp{s_{n-1}}.
\ee

This equation can be seen as a generalization of the usual
holomorphic anomaly equation. If we only put the cubic coupling
$s_2=-i\l$ to a non-zero value (the string coupling constant)
then equation (\ref{hol-ano}) reduced to (with $s_1=-i\tau$)
\be
{\d Z \over \d\taubar}={\l^2\over \t_2^2} {\d^2 Z\over \d\tau^2},
\label{anomaly}
\ee
which is of the form given in \cite{bcov1}.

\newsubsection{The two-dimensional QCD string revisited}
 
Let us now finally return to our original motivation and reconsider the
implications of all this for the $(\d\v)^3$ model and the 2d QCD string. 
The expression for the torus partition function $Z(\tau,\l)$ in terms
of the string coupling constant $\l=1/N$ as it follows from the
large $N$ expansion of the Yang-Mills partition function is 
given by \cite{douglas}
(see also \cite{texel} for a short derivation using branched covers)
\be
Z(\tau,\l)=\oint {dy\over 2\pi i y}
\prod_{p\in\Z_{\geq 0}+{1\over 2}}
\left(1+y q^p e^{\lambda p^2/ 2}\right)
\left(1+y\inv q^p e^{-\lambda p^2/ 2}\right)
\ee
We recognize the QCD string partition function as a generalized
$\woinf$ character where we added the spin three interaction term
\be
\oint \F^2 = - \oint b\d^2c = \oint {i\over 3}(\d\v)^3.
\ee
and evaluate in the zero $U(1)$ charge sector.

One of the interesting properties of the QCD string partition function
is that the coefficients $F_g(\tau)$ in the perturbative string
expansion
\be
Z(\tau,\l)=\exp\sum_{g\geq 1} \l^{2g-2} F_g(\tau)
\ee
have rather intricated modular properties. 
They are so-called quasi-modular forms, of weight $6g-6$ 
\cite{rudd,texel,zagier}. Quasi-modular forms are polynomials in the 
Eisenstein series $E_2,E_4,E_6$. The Eisenstein series $E_4$ and $E_6$
are modular forms of weight $4$ and $6$, and generate the ring of all
modular forms. The form $E_2$ is not quite modular of weight two, but
has a modular anomaly
\be
E_2\left({a\tau+b\over c\tau +d}\right) = (c\tau+d)^2 E_2(\tau)
+ {12\over 2\pi i} c(c\tau+d).
\ee
However, $E_2$ can be made into a proper modular form by adding an
anholomorphic term and defining $E_2^*$ as in (\ref{Etwo}). That is to
say, the string partition function should be regarded as the limit
$\taubar \ra -i\infty$ of an expression that is no longer holomorphic
in $\tau$ but that is modular invariant, if one let $\l$ transform
with modular weight $-3$.  In fact, this suggests that the ``correct''
string field theory lagrangian is given by
\be 
S= \int {d^2z\over \pi} \left({1\over 2}
\d \v\dbar\v + {\l\over 6} (-i\d\v)^3\right).
\ee
or equivalently
\be
S= \int {d^2z\over \pi} \left(b\dbar c + \l\, b \d^2 c\right).
\ee
This is in complete accordance with the philosophy of
\cite{bcov1,bcov2}, where it was shown that the (topological) string
on a Calabi-Yau manifold has anti-holomorphic dependence. Only if we
decouple the anti-holomorphic couplings, do we recover the
(holomorphic) instanton counting functions.  The fact that the string
coupling has modular weight $-3$ is also consistent with this point of
view.

Since we only recover the QCD answer in the $\taubar\ra -i\infty$
limit, the above action is not uniquely determined. In fact, if our
starting point was a pure cubic Hamiltonian, then according to our
general formulas, the corresponding Lagrangian would contain higher
order terms corresponding to $(\d\v)^n$ interactions with arbitrary
$n>3$. The corresponding couplings would however go to zero in the
holomorphic limit.

This is important if we want to make contact with the holomorphic
anomaly equation. Indeed, only with a cubic Hamiltonian (and thus a
non-polynomial Lagrangian) do we recover the simple form
(\ref{anomaly}). Vice versa, a cubic Lagrangian will give a
non-polynomial Hamiltonian which satisfies an anomaly equation of
general type (\ref{hol-ano}). So we find that the two characteristic
features of the six-dimensional Kodaira-Spencer theory --- a simple
cubic $(\d\v)^3$ Lagrangian and a simple quadratic holomorphic anomaly
equation --- are incompatible in two-dimensions. An open question is
whether the more general holomorphic anomaly equation that we derived
in \S4.6 also occurs in the superstring context.

\bigskip
\bigskip

{\noindent \sc Acknowledgements} 

I would like to thank C.\ Vafa for collaboration in the initial phase
of this work and acknowledge useful discussions with D. Bernard,
M. Douglas, V. Kac, O. de Mirleau, G. Moore, D. Neumann and D. Zagier.

\renewcommand{\Large}{\normalsize} 

\end{document}